\documentclass{aa}
\usepackage{txfonts}
\usepackage{times,psfig}

\def\ltsima{$\; \buildrel < \over \sim \;$}
\def\lsim{\lower.5ex\hbox{\ltsima}}
\def\gtsima{$\; \buildrel > \over \sim \;$}
\def\gsim{\lower.5ex\hbox{\gtsima}}

\newcommand{\be}{\begin{equation}}
\newcommand{\en}{\end{equation}}
\newcommand{\ergs}{\rm \ erg \; s^{-1}}

\def\rsole {~R_{\odot}}
\def\msole {~M_{\odot}}

\begin{document}

\title{Doppler tomography of the transient X-ray binary Centaurus X-4 in
quiescence\thanks{The results reported in this paper are partially based on 
observations carried out at ESO, La Silla, Chile (67.D-0116).}}


\author{P. D'Avanzo\inst{1,2}, S.~Campana\inst{1},  
J. Casares\inst{3}, G.L. Israel\inst{4},
S. Covino\inst{1}, P.A. Charles\inst{5,6} 
L. Stella\inst{4}}   

\authorrunning{D'Avanzo et al.}

\titlerunning{Doppler tomography of Cen X-4}

\offprints{Paolo D'Avanzo, davanzo@merate.mi.astro.it}

\institute{INAF-Osservatorio Astronomico di Brera, Via Bianchi 46, I--23807
Merate (Lc), Italy
\and
Universit\`a degli Studi dell'Insubria, Dipartimento di Fisica e Matematica, Via Valleggio
11, I--22100 Como, Italy
\and
Instituto de Astrof\`{\i}sica de Canarias, 38200 La Laguna,
Tenerife, Spain
\and
INAF-Osservatorio Astronomico di Roma,
Via Frascati 33, I--00040 Monteporzio Catone (Roma), Italy
\and
School of Physics \& Astronomy, University of
Southampton, Southampton SO17 1BJ, UK
\and
South African Astronomical Observatory, P O Box 9, Observatory 7935, South Africa}

\date{Received; accepted}

\abstract{ We present ESO-NTT low resolution spectroscopy of the
neutron star X-ray transient Cen X-4 in quiescence over a complete
orbital cycle. Our data reveal the presence of a K3-7 V companion
which contributes $63\%$ to the 5600--6900~$\AA$ flux and orbits the
neutron star with a velocity semi-amplitude of $K_2=145.8\pm 1.0$ km
s$^{-1}$. This, combined with a previous determination of the
inclination angle and mass ratio, yields a neutron star and companion
mass of $M_1=1.5\pm1.0\msole$ and $M_2=0.31\pm0.27\msole$,
respectively. The mass donor is thus undermassive for the inferred
spectral type indicating it is probably evolved, in agreement with
previous studies. Doppler tomography of the H${\alpha}$ line shows
prominent emission located on the companion and a slightly asymmetric
accretion disc distribution similar to that seen in systems with
precessing eccentric discs. Strong H${\alpha}$ emission from the
companion can be explained by X-ray irradiation from the primary. No
evidence is found for a hot spot in H${\alpha}$, whereas one is
revealed via Doppler tomography of the HeI lines. This can be
interpreted as the hot spot and outer regions of the disc being at a
higher temperature than in other systems.


}

\maketitle

\section{Introduction}

In recent years coherent, fast pulsations have been discovered in
six Low Mass X-ray Binaries (LMXBs) leading to the conclusion that they
contain rapidly spinning ($\sim 2-5$ ms), weakly magnetic neutron
stars (see e.g. Chakrabarty 2004).
X-ray light curves obtained with RXTE also revealed coherent oscillations
during Type I X-ray bursts (i.e. thermonuclear explosions on the
neutron star surface) in fourteen LMXBs (e.g. van der Klis
2000). These findings directly confirmed evolutionary models that link
the LMXB neutron stars to those of millisecond radio pulsars, the
former being the progenitors of the latter (e.g. Bhattacharya \& van
den Heuvel 1991).

A large fraction of LMXBs accrete matter at very high rates (and
therefore shine as bright X-ray sources) only sporadically. Among
these systems are Soft X-ray Transients (SXTs), hosting old neutron
stars (for a review see Campana et al. 1998). These systems alternate
periods (weeks to months) of high X-ray luminosity with long (1--10
years) intervals of quiescence in which the X-ray luminosity drops by
some 5--6 orders of magnitude. Concerning the optical properties, SXTs
in outburst are relatively bright and their emission is dominated by
X-ray reprocessing in the accretion disc surrounding the neutron
star. In quiescence their optical luminosity drops by as much as 6--7
mag, thereby allowing the late type companion to be detected and
studied. Weak, residual optical emission is still present in
quiescence, usually at the $\sim 10-30\%$ level in V, that is usually
ascribed to a dim, viscously~-~heated accretion disc (van Paradijs \&
McClintock 1995).

The large luminosity variations of SXTs correspond to vast changes in
the accretion rate onto the collapsed objects. These provide an
opportunity to explore a range of accretion regimes onto the neutron
star that are inaccessible to persistent sources. While in outburst
accretion onto the neutron star surface takes place in a manner
similar to that of persistent LMXBs, whereas when the accretion rate
decreases, matter is halted at the magnetospheric boundary by a
centrifugal barrier generated by the fast rotating neutron star
magnetosphere (a.k.a. propeller effect; Illarionov \& Sunyaev
1975). If the mass inflow rate decreases further, the neutron star can
enter the radio pulsar regime and begin to expel the inflowing
material from the companion star (Shaham \& Tavani 1991; Stella et
al. 1994; Campana et al. 1995; Burderi et al. 2001).

Some progress has recently been made in the study of the outburst
decay and quiescent state of some of these systems (Asai et al. 1996),
yet the interpretation of the quiescent optical and X-ray observations
is ambiguous. The quiescent X-ray spectrum is well fit by the sum of a
soft component (usually a neutron star atmosphere model with
temperature $k\,T\sim 0.1-0.3$ keV) plus a hard tail power law with
photon index in the range 1--2. There are cases in which only one of
the two components is detected.  Different interpretations have been
put forward. Menou et al. (1999) suggested the presence of an
Advection Dominated Accretion Flow (ADAF) dominating the residual
optical emission: some of the inflowing matter would leak through the
magnetosphere and accrete onto the neutron star surface, giving rise
to the soft X-ray component. Campana et al. (1998a, b, see also
Campana \& Stella 2000) suggested instead that the soft component
arises from the cooling of the neutron star surface heated during the
repeated outburst episodes (see also Brown et al. 1998; Rutledge et
al. 1999) and the hard tail arises from shock emission powered by an
active radio pulsar. In this case (part of) the residual quiescent
optical flux may come from the shock emission in the interaction
between the radio pulsar wind and the matter inflowing from the
companion. 

We therefore decided to utilise Doppler tomography in order to study
the kinematics and geometry of matter in such interacting
binaries.  This would allow us to derive two-dimensional velocity maps
of the optical line-emitting regions, by combining spectra taken at
different orbital phases (Marsh \& Horne 1988).  This method provided
strong evidence for the presence of accretion discs in cataclysmic
variables (CVs) and quiescent black hole binaries (e.g. Casares et
al. 1997). In the case of the CV AE Aqr, Doppler tomograms revealed a
highly unusual configuration, which supported the idea that the white
dwarf expels most of the incoming material from the companion star and
is thus in the propeller regime (Wynn et al. 1997; Horne 1999). The
potential of this technique has not yet been fully exploited for
neutron star SXTs. Only recently has a Doppler map of the SXT Cen X-4
been obtained (Torres et al. 2002), although this required
observations acquired over a year. In addition, Doppler maps of XTE
J2123-058 in both outburst and quiescence have been produced by Hynes
et al. (2001) and Casares et al. (2002), respectively.

\section{Cen X-4}

Bright X-ray outbursts were detected from Cen X-4 in 1969 (Conner et
al. 1969) and again in 1979 (Kaluzienski et al. 1980), since when it
has been quiescent.  The companion star is an evolved $\sim
0.7\msole$, K5--7 star (Shahbaz et al. 1993; Torres et al. 2002)
making Cen X-4 one of the brightest SXTs in quiescence with V=18.7
and (A$_{\rm V}=0.3$ mag). The stellar parameters have recently been
improved by Gonzalez Hernandez et al. (2005) who found a surface
temperature of T$_{eff}=4500 \pm 100$ K and surface gravity of
log$g=3.9 \pm 0.3$. A 15.1 hr orbital period was determined from the
sinusoidal variation of the optical light curve (Cowley et al. 1988;
Chevalier et al. 1989; McClintock \& Remillard 1990).  These
characteristics make Cen X-4 an excellent candidate for Doppler
tomography studies in quiescence.

The optical spectrum shows clear residual disk emission after
subtraction of the companion's spectrum. Assuming a smooth spectral
energy distribution, it contributes $80\%$ in B, $30\%$ in V, $25\%$
in R and $10\%$ in I (Shahbaz, Naylor \& Charles 1993). Strong
H$\alpha$ ($EW\sim 40$ $\AA$), H$\beta$ ($EW\sim 20$ $\AA$) and
$H\gamma$ ($EW\sim 10$ $\AA$) emission is also present. Although variable,
these lines are always present in the optical spectra taken at
different epochs.  For instance, van Paradijs et al. (1989) reported
equivalent widths of 43, 33 and 28 $\AA$, respectively.

HST observations revealed a flat UV spectrum (in a $\nu F_{\nu}$
vs. $\nu$ representation McClintock \& Remillard 2000), which
connected smoothly with the hard tail X-ray power law (e.g. Campana \&
Stella 2000), suggestive of a shock spectrum. This is at variance with
observations of the black hole candidate A0620--00.

\section{Observational data and reduction}

Cen X-4 spectra were taken on 26-28 May 2001 with the ESO 3.5m New
Technology Telescope (NTT) at La Silla. We used the ESO Multi Mode
Instrument (EMMI) with a Tektronix CCD of $2048\times2048$ pixels
yielding a field of view of $9.1\arcmin\times 8.6\arcmin$ at a
resolution of 0.27$\arcsec$/pixel.  All nights were photometric with
seeing in the $0.5-1\arcsec$ range.  We used a $1\arcsec$ slit and
grating 7 to cover the H$\alpha$ region, centered at 6250 $\AA$, and
with a dispersion of 0.66 $\AA$/pix.

The data were reduced using standard {\sc ESO-MIDAS} procedures for
bias subtraction, flat-field correction and cosmic ray removal. All
spectra were sky-subtracted and corrected for atmospheric
extinction. Helium-argon lamp images were obtained for wavelength
calibration during daytime and with the telescope vertically parked.
The wavelength scale was then derived through third-order polynomial
fits to 26 lines, resulting in an rms scatter of $<0.06$
$\AA$. Instrumental flexure during our observations were then accounted for using
atmospheric emission lines in the sky spectra. Spectral resolution was about
$\sim 2.2$ pixels, or $\sim 70$ km s$^{-1}$ in velocity.

Observations were carried out over three consecutive nights so as to
completely cover the 15.1 hr orbital cycle of Cen X-4. We obtained 30
spectra, each 30 min long.  This is the first time that an extensive
and contiguous orbital coverage has been obtained for Cen X-4.

\begin{table}
\caption{Radial velocity parameters of Cen X-4.}
\begin{tabular}{ccccc}
\hline
Template Star &  Spectral Type &   $\gamma$     &   $K_2$       &${\chi^2}_{\rm red}$\\
         &                & (km s$^{-1}$)  & (km s$^{-1}$) &\\ 
\hline
HR 5536    &  K3 III      & $203.6\pm 2.1$  & $147.7\pm 1.0$&1.8\\
HD 184467  &  K2 V        & $191.2\pm 2.1$  & $147.6\pm 0.9$&1.9\\
HD 29697   &  K3 V        & $209  \pm 10 $  & $146.1\pm 1.0$&1.2\\
HD 154712A &  K4 V        & $194.2\pm 2.1$  & $146.2\pm 0.8$&1.7\\
61 Cyg A   &  K5 V        & $196.4\pm 1.0$  & $146.4\pm 0.8$&1.9\\
61 Cyg B   &  K7 V        & $196.0\pm 1.1$  & $145.8\pm 1.0$&1.1\\
\hline
\end{tabular}
\label{cross}
\end{table}
\begin{table*}
\begin{center}
\caption{Cen X-4 dynamical parameters.}
\begin{tabular}{cccc}
\hline
Reference                   &   $\gamma$      & $K_2$           & $P$ \\
                            &  (km s$^{-1}$)  & (km s$^{-1}$)   &   (d)       \\
\hline
Cowley et al. 1988          & $234\pm 8$      & $152\pm 10$     &    -     \\
McClintock \& Remillard 1990& $137\pm 17$     & $146\pm 12$     & 0.629063 \\
Torres et al. 2002          & $184.0\pm 1.9$  & $150.0\pm 1.1$  & $0.6290496\pm 0.0000021$\\ 
This paper                  & $196.0\pm 1.1$  & $145.8\pm1.0$   & $0.630\pm0.001$ \\
\hline
\end{tabular}
\label{param}
\end{center}
\footnotesize{The velocity values reported
here have been obtained by fixing $P$ to the value given by Torres et
al. (2002). We also repeated the fit leaving all the parameters free,
obtaining the value given here.} 
\end{table*}

\section{Cen X-4 properties}
 
\subsection{Radial velocity curve}

Radial velocities were measured from our 30 spectra through
cross-correlation with late-type template stars. Prior to cross-correlating, the
spectra were rebinned to a uniform velocity scale. Cross-correlation
was performed in the range 5930-6750 $\AA$, after masking out strong
night-sky lines due to OI 6300 $\AA$, the main atmospheric and
interstellar features, emission lines from the accretion disc
(H$\alpha$, HeI $\lambda$5876 and $\lambda$6678) and our spectral
extremes, resulting in an effective wavelength range of 6380--6520
$\AA$. We then performed a (least-squares) sine fit to our velocity
data. We adopted the parameters obtained with the K7 V template (see
Table \ref{cross}), as they provided the best fit in terms of reduced
$\chi^2$, giving $K_2 = 145.8 \pm 1.0$ km s$^{-1}$, and
${\gamma}=196.0\pm 1.1$ km s$^{-1}$, which are, respectively, the
radial velocity amplitude and systemic velocity of Cen X-4.

The dynamical parameters obtained here are given in Table \ref{param},
together with the results of Cowley et al. (1988), McClintock \&
Remillard (1990) and Torres et al. (2002). Our value for $K_2$ is
slightly smaller than that of Torres et al. (2002) and is
statistically inconsistent. However, we note that the data of Torres
et al. are spread over several years, whereas ours were
collected over three consecutive nights. Our result is instead
consistent with those of Cowley et al. (1988) and McClintock \&
Remillard (1990). Curiously, a much larger variation is found in the
systemic velocity $\gamma$, as has also been noted by Torres et
al. (2002). We obtain a $\gamma$ intermediate to the values reported
earlier. We do not have an explanation for this discrepancy, which in
principle could be due to the use of an erroneous heliocentric radial
velocity for the template stars. Our value for $P$ has a correspondingly
large uncertainty as a result of our relatively short observing
baseline with respect to, e.g. Torres et al. (2002).

\begin{figure*}[htbp]
\begin{center}
\psfig{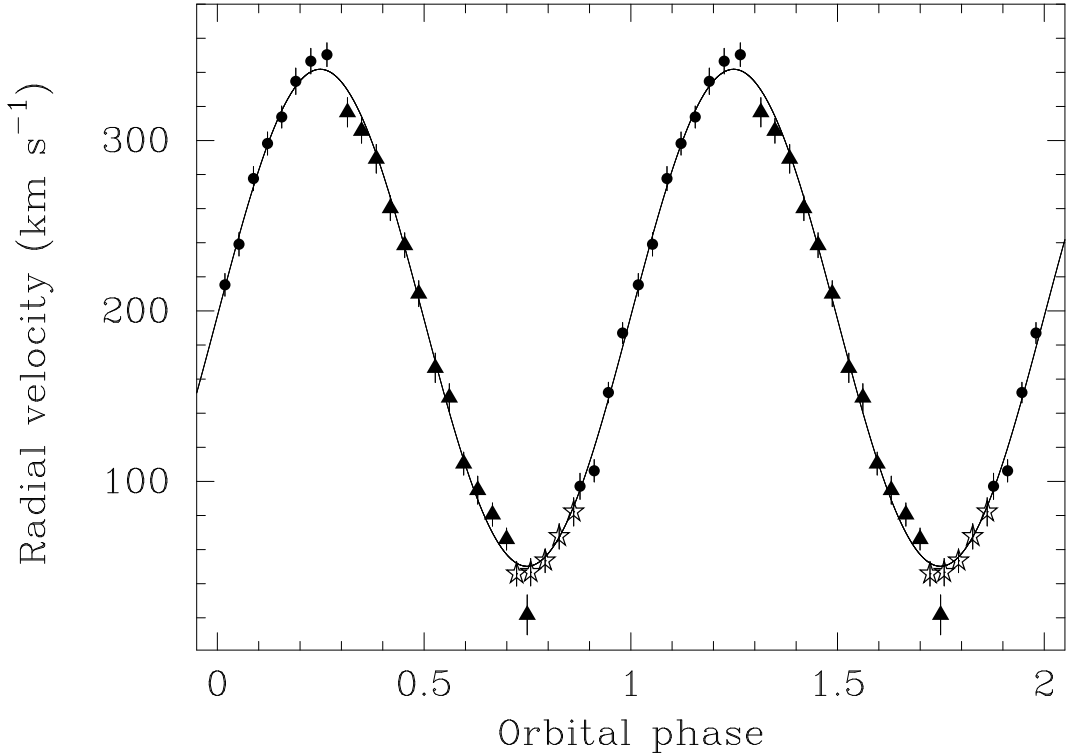}
\end{center}
\caption{Radial velocity curve for Cen X-4 computed using a K7 V template. Observations were performed on 26 (triangles), 27 (dots) and 28 (stars) May 2001. Two phases are shown for clarity. The best sine-wave fit is also shown (see Section 4.1).}
\label{curve}
\end{figure*}

The radial velocity curve, after folding with our best ephemeris, is
shown in Fig. \ref{curve}. It is highly sinusoidal
(${\chi^2}_{\nu}=1.1$) despite the point at phase 0.75. This
point is not due to instrumental problems (such as telescope flexure)
and we also note that a similar effect is seen in the radial velocity
curve of Cen X-4 measured by Torres et al. (2002); however there is
no clear explanation for such behaviour.

\subsection{Spectral classification}

\begin{figure*}[htbp]
\begin{center}
\psfig{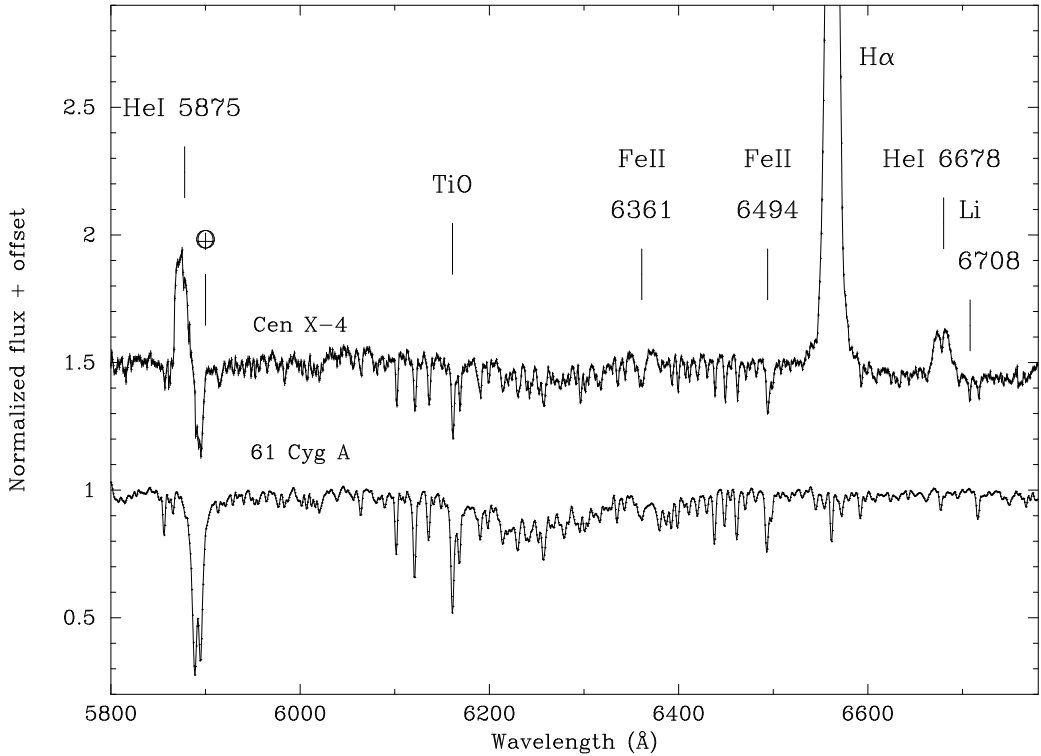}
\end{center}
\caption{Doppler-corrected average of the 30 spectra of Cen X-4 in quiescence that were obtained and
analyzed in the present study (top spectrum) compared  with a K5 V template (61 Cyg A). Arbitrary vertical offsets
have been applied for clarity.}
\label{tot_spectrum}
\end{figure*}

We have produced an average spectrum of Cen X-4 in the rest frame of
the companion star after Doppler shifting every individual spectrum
using our orbital solution and the orbital period of Torres et
al. (2002). The spectrum shows prominent H${\alpha}$, HeI
$\lambda$5876, $\lambda$6678 emission as well as absorption features
typical of late type stars (e.g. TiO $\lambda$6161, FeII $\lambda$6361
and FeII $\lambda$6494, Fig. \ref{tot_spectrum}). A spectral
classification can be derived by taking different templates degraded
appropriately with a Gaussian bandpass to match the broadening of our
NTT observations. These templates were then multiplied by a factor $0
\leq f \leq 1$, representing the fractional contribution of light from
the secondary star and subtracted from the target average (after
masking interstellar, night-sky and emission lines and rebinning to a
uniform velocity scale). The best fit is obtained for templates K3
V--K5 V and $f=0.59-0.62$, although a K7 III also provides an
acceptable fit (see Table \ref{template}). More importantly, in the
Doppler-corrected average, Li I $\lambda$6708 absorption becomes prominent. The
detection of a strong Li $\lambda$6708 absorption feature in such a late-type
companion would normally be unexpected since the star's initial Li I
content should be rapidly depleted by a contamination of convective
mixing and mass transfer to the compact object.  Such unusual abundances
of Li I have already been found by other authors (e.g. Martin et
al. 1994, Torres et al. 2002). This significant and anomalous
abundance of Li I in Cen X-4 (and other transients) finds various
explanations in the literature, such as synthesis in a supernova
explosion of the compact primary's progenitor, or $\alpha$-$\alpha$
reactions during the repeated strong outbursts that characterize
transient X-ray binaries (Martin et al. 1994), a relatively recent
formation of the system (Gonzalez Hernandez et al. 2005), or an effect
related to the tidally-locked rotation of the two stars, which leads to
a slower lithium destruction rate (Maccarone et al. 2005).

An alternative approach to the spectral classification consists of
comparing the ratio of the equivalent width of the absorption lines in
Cen X-4 with different templates. We consider the ratios
$\lambda$6361(FeII)/$\lambda$6161 (TiO) and
$\lambda$6494(FeII)/$\lambda$6161 (TiO), which also classify the
secondary of Cen X-4 as a K5 V--K7 V star (see Table \ref{ratio}). We
therefore conclude that the most likely spectral type for Cen X-4 is a
K3--7 V star.

\begin{table}
\caption{Cen X-4 spectral classification by direct fitting}
\begin{tabular}{cccccc}
\hline
Template &  Spectral Type &  $f$    &${\chi^2}_{\nu}$\\
         &                &         &(117 d.o.f.)\\ 
\hline
HR 5536    &  K3 III	  &$0.49\pm0.01$ & 2.29 \\
HR 6159    &  K4 III	  &$0.44\pm0.01$ & 2.40 \\
HR 5622    &  K5 III	  &$0.53\pm0.01$ & 2.30 \\
HD 184467  &  K7 III	  &$0.84\pm0.05$ & 1.30 \\
HD 29697   &  K3 V	  &$0.62\pm0.04$ & 1.29 \\
HD 154712A &  K4 V	  &$0.65\pm0.04$ & 1.32 \\
61 Cyg A   &  K5 V	  &$0.63\pm0.04$ & 1.30 \\
61 Cyg B   &  K7 V	  &$0.59\pm0.04$ & 1.39 \\
\hline
\end{tabular}
\label{template}
\end{table}

\begin{table}
\caption{Line ratios for Cen X-4 and template stars.}
\begin{tabular}{cccc}
\hline
Template   &  Spectral Type  & 6361(FeII)/    & 6494(FeII)/ \\
	   &                 & 6161 (TiO)     & 6161 (TiO)  \\
\hline		
Cen X-4    &                 & $0.41\pm0.06$  &$0.52\pm0.06$\\
\hline
HR 5536    &  K3 III         & 1.43	      &1.52 \\
HR 6159    &  K4 III         & 1.59	      &1.88 \\
HR 5622    &  K5 III         & 0.76	      &0.86 \\
HD 184467  &  K7 III         & 0.68	      &0.78 \\
HD 29697   &  K3 V           & 0.18	      &0.44 \\
HD 154712A &  K4 V           & 0.12	      &0.51 \\
61 Cyg A   &  K5 V           & 0.33	      &0.52 \\
61 Cyg B   &  K7 V           & 0.50	      &0.56 \\
\hline
\end{tabular}
\label{ratio}
\end{table}

\subsection{Orbital variation of the companion star contribution}

\begin{figure*}
\begin{center}
\psfig{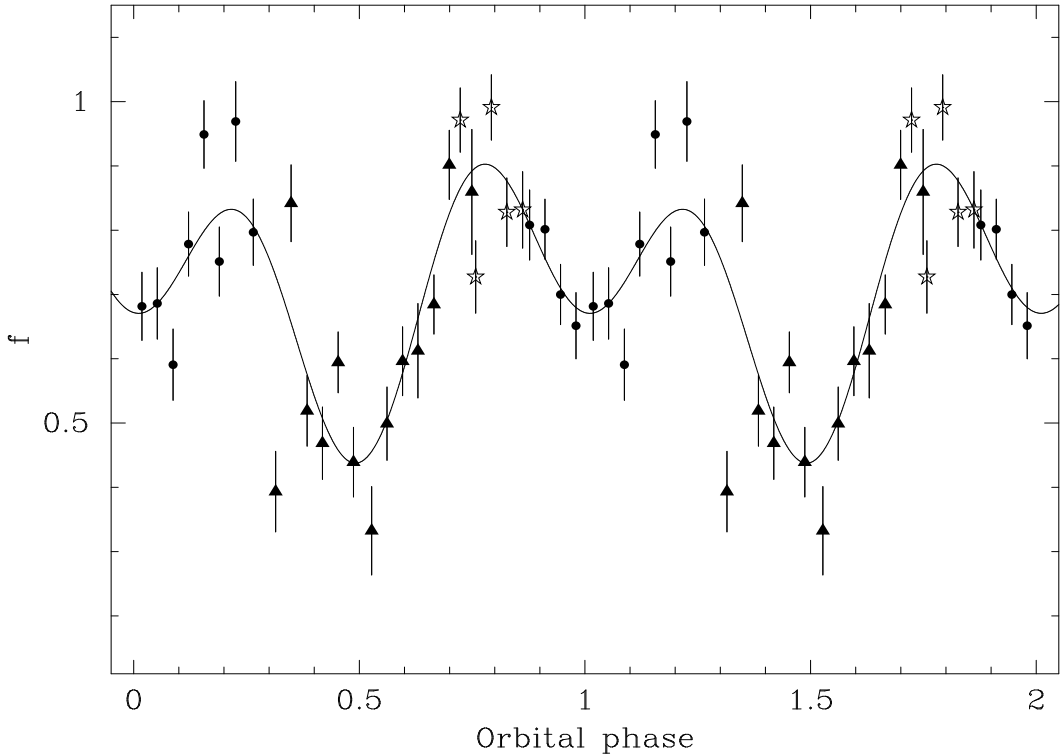}
\caption{Variation of fractional contribution $f$ of the companion star with orbital phase. Observations were performed on 26 (triangles), 27 (dots) and 28 (stars) May 2001. Two cycles are shown for clarity.}
\label{fvar}
\end{center}
\end{figure*}

Taking advantage of the good signal to noise of each individual
spectrum we tried to estimate the factor $f$ for all the spectra,
allowing us to monitor its variability with orbital phase. It is interesting
to note that $f$ is not constant but is modulated at the orbital
period (see Fig. \ref{fvar}). Variations of $f$ with phase show two
minima at phase 0 (i.e. at superior conjunction) and 0.5 and two
nearly equal maxima at phase 0.25 and 0.75. The latter occur when the
observer sees the elongated star with maximum projected surface
area. This behaviour follows the classical ellipsoidal modulation
previously observed in optical and IR lightcurves (e.g. McClintock \&
Remillard 1990; Shahbaz et al. 1993), and strongly suggests that it is
dominated by the visibility of the tidally distorted
companion. However, we note that gravity darkening effects, expected
from the classical ellipsoidal modulation, are not sufficient to
explain the large depth difference between phase 0 and 0.5 observed in
our lightcurve. Such a large difference may require the presence of an
additional effect, such as the heating of the inner hemisphere of the
companion by X-ray irradiation from the compact object. We give
further evidence of such a phenomenon in Section 5, in the light of
our Doppler tomography results.

\subsection{System parameters}

If we assume that the secondary star is syncronized with the binary
motion and fills its Roche lobe, we can determine the mass ratio ($q =
M_2/M_1$) through the expression (see e.g. Wade \& Horne 1988)

$v$\, $\sin i = 0.462K_2 q^{1/3}(1 + q)^{2/3}$.

Taking the secondary star's rotational broadening as measured by
Torres et al. (2002) ($v$\, $\sin i = 43 \pm 6$ km s$^{-1}$), and
combined with our value of $K_2$, we derive $q = 0.18 \pm
0.06$. Combining our value of $K_2$ with the precise orbital period
reported by Torres et al. (2002), we derive a mass function of $f(M) =
0.201 \pm 0.004 \msole$. In addition, if we take again the $v$\, $\sin
i$ value measured by Torres et al. (2002) and the constraint on $i$
from the IR observations of Shahbaz et al. (1993, i.e. $31^{\circ} < i
< 54^{\circ}$ at $90\%$ confidence level), we can constrain the masses
of the two stars: $0.47 < M_{NS} < 2.31 \msole$ for the neutron star
and $0.06 < M_2 < 0.55$ for the companion. $M_2$ is clearly
undermassive for the inferred spectral type as discussed by Shahbaz et
al.


\section{Doppler tomography}

\begin{figure*}
\begin{center}
\psfig{figure=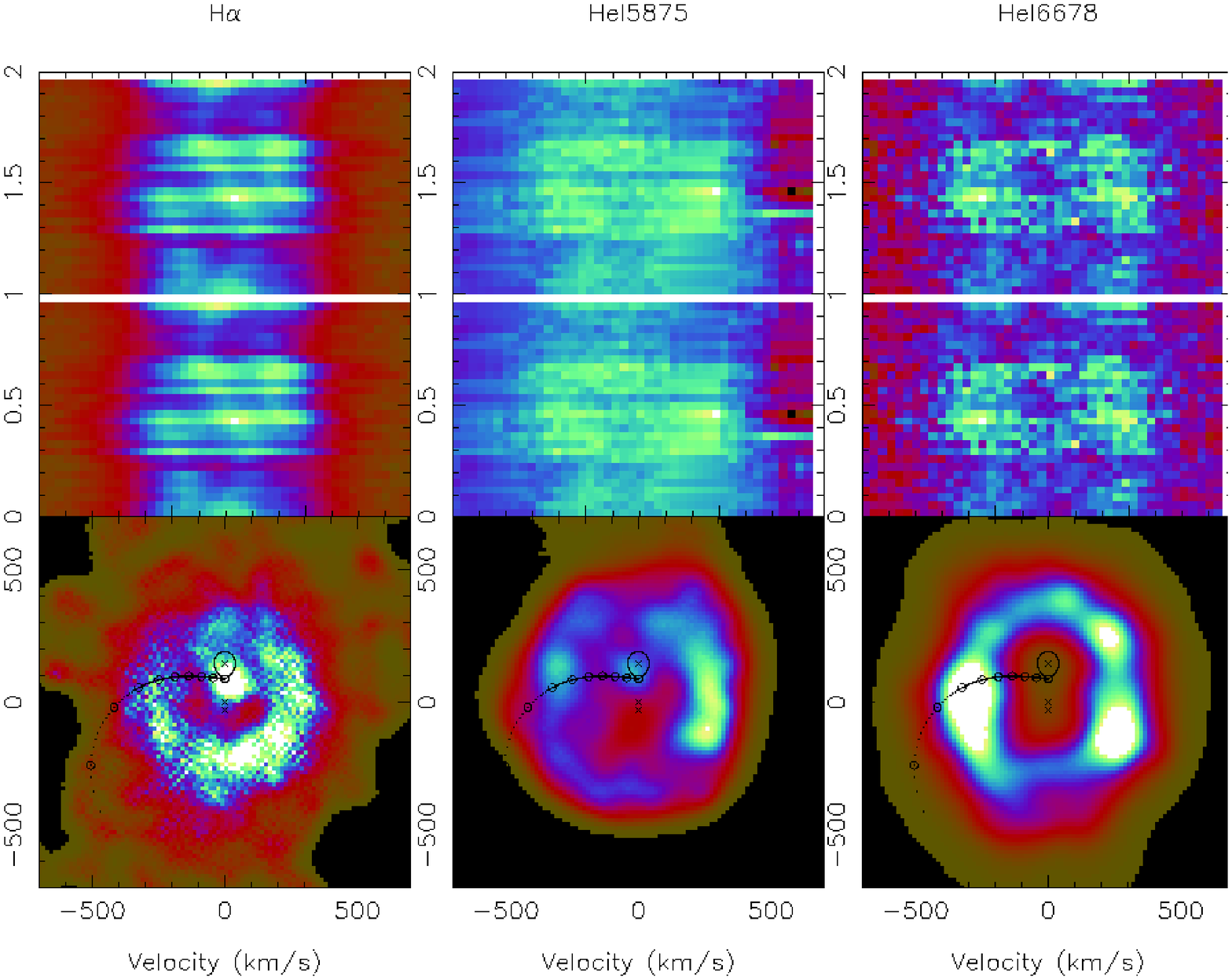,height=14cm,angle=0}
\caption{Doppler maps of the H$\alpha$, HeI $\lambda$5876 and HeI $\lambda$6678
lines. The trailed spectra are shown in the upper panels and, for clarity, the
same cycle has been plotted twice. The Roche lobe of the secondary and the predicted 
velocity of the gas stream are plotted for $K_2 = 145.8$ km s$^{-1}$ and $q=0.18$. Distances are marked along the curve in open circles and are in multiples of 0.1$R_{L1}$. The center of mass of the system and the neutron star position are denoted by crosses.}
\label{doppler}
\end{center}
\end{figure*}

The three emission lines seen in the Cen X-4 spectra are double-peaked,
testifying to the presence of an axisymmetric emission region around the compact
object. The line profiles vary with orbital phase and there is evidence for 
different emitting components.
Algorithms have been developed to reconstruct the velocity field of
the line emitting region, i.e. the Doppler tomography technique (Marsh
\& Horne 1988).  The continua were normalized and subtracted so as to
give a pure emission line profile which was subsequently rebinned onto
a constant velocity scale of 30.5 km s$^{-1}$ pixel$^{-1}$. The
resulting Doppler maps are presented in Fig. \ref{doppler} in velocity
coordinates, together with the trailed spectra. The Roche lobe of the
companion and the theoretical path of the gas stream for a $K_2=145.8$
km s$^{-1}$ and $q=0.18$ system are superposed over each map, in steps
of $0.1$ R$_{L_1}$.

\subsection{H$\alpha$ tomography} 

The H$\alpha$ emission line has a mean equivalent width (EW) of $45.3
\pm 1.2$ $\AA$.  The H$\alpha$ map is dominated by a remarkably
intense spot, coincident in phase and velocity with the companion
star. This may indicate either chromospheric activity in the
synchronized (and hence rapidly rotating) late-type companion or
irradiation from the compact object. There is also evidence for a
ring-like structure, a signature of a classic accretion disc. We note
that the brightness distribution of the disc is slightly asymmetric,
with enhanced emission at the negative $V_y$ quadrants. This is
somewhat reminiscent of the emissivity distribution observed in SW Sex
stars, a group of persistent cataclysmic variables where exotic
scenarios (namely magnetic accretion, disc overflow and propeller
mechanisms) have been invoked to explain their peculiar emission line
profiles. The same phenomenology is sometimes observed in LMXBs both
during outburst and quiescence (e.g. XTE J2123-058: Hynes et al. 2001;
Casares et al. 2002). We note, however, that one of the signatures of
SW Sex stars is the presence of low-velocity emission in their Doppler
maps (e.g. Wynn et al. 1997; Welsh, Horne \& Gomer 1998). Although our
H$\alpha$ tomogram shows an asymmetric distribution, there is no
evidence for emitting structures at velocities lower than $\sim$200 km
s$^{-1}$ and hence a SW Sex scenario does not seems a likely
model. Alternatively, asymmetric disc emissivity can be produced by
spiral waves (triggered by tidal interaction with the companion star,
see Steeghs, Harlaftis \& Horne 1998) or superhumps caused by
eccentric precessing discs (Foulkes et al. 2004). Finally, direct
irradiation of the outer disc rim can also excite enhanced
emission. In this latter case, our asymmetric H$\alpha$ tomogram may
indicate a vertical irradiated structure around orbital phases
0.3-0.5, possibly connected with a shock front.

On the other hand, we see no sign of emission from the expected hot spot
location. This pattern has also been observed in other quiescent (black hole) 
X-ray transients, i.e. Nova Mus. 91 (Casares et al. 1997), Nova Oph. 77 
(Harlaftis et al 1997), XTE J1118+480 (Torres et al. 2005).  

In order to further investigate the possible origins of the H$\alpha$
emission coming from the secondary, we extracted the narrow H$\alpha$
component from our Doppler map. To do this we subtracted from our map
a simulated axisymmetric image centered on the compact object ($V_x =
0, V_y = - K_2/q$), obtaining a map where the emission arising from
the secondary becomes clearer (Fig. ref{dopsub}, first panel). Then,
projecting this new Doppler image into the observed orbital phases and
adding all the residuals together we obtained a spectrum which
represents the H$\alpha$ emission from the secondary. Following
Casares et al. (1997) and Torres et al. (2002) we estimated an
equivalent width for the narrow line component of H$\alpha$ of $4.4
\pm 0.5$ $\AA$, taking into account the veiling from the accretion
disc. This is quite a high value of H$\alpha$ emission for such
systems. Using the relationship given by Soderblom et al. (1993):
$$
log\,F_{H_{\alpha}} = log\,EW(H_{\alpha})\,+\,0.113\,(B-V)^2_0\,-\,1.188\,(B-V)_0\,+7.487
$$
\noindent we convert this value into an H$\alpha$ line surface flux at
the star ($F_{H\alpha}$). Using $(B-V)_{0}=1.25$ mag (Chevalier et
al. 1989) we obtain $F_{H\alpha}=7\times10^6$ erg cm$^{-2}$ s$^{-1}$.

A chromospherically active secondary is a possible explanation for
this H$\alpha$ emission. We can compare our estimates with the results
reported by Soderblom et al. (1993) for rapidly rotating dwarfs in the
Pleiades. From Table 6 of that paper we find that the
H$\alpha$ emission of the Cen X-4 companion is higher in strength than
that observed for all the K dwarfs in their sample. Moreover, we note
that Torres et al. (2002) found for the narrow line component of
H$\alpha$ an EW of $\sim 2.3$\AA and a veiling factor ($1-f$) of 0.25
(where $f$ is the fractional contribution of light from the secondary
star) while our higher value of EW corresponds to a higher value of
($1-f$) (0.37, see Table \ref{template}), showing that the veiling and
H$\alpha$ EW are correlated.  This is not expected in a chromospheric
H$\alpha$ emission scenario.

X-ray irradiation by the compact object is an alternative explanation
for the H$\alpha$ emission from the secondary. The intrinsic X-ray
luminosity in the 0.5-10 keV band was estimated by Campana et
al. (2004) to be $L \simeq 4\times 10^{32}\ergs$. The X-ray
irradiation at the secondary star is then $F{_X}=L{_X}/(4\pi
a{^2})=5\times10^{8}$ erg cm$^{-2}$ s$^{-1}$, for an orbital
separation $a=3.6\rsole$, that can power the H$\alpha$ emission if at
least 1\% of the incident X-ray flux is reprocessed to H$\alpha$
photons.

Following Hynes et al. (2002), a rough estimate of the fraction of the
X-ray luminosity converted to H$\alpha$ photons can be obtained
assuming that the observed H$\alpha$ emission is due completely
to X-ray reprocessing by the companion:
$$
L_{H\alpha}=f_1f_2L_X
$$ 
where $f_1$ is the fraction of the high-energy emission $L_X$ which
is intercepted by the companion (i.e. the solid angle of the
companion) and $f_2$ is the fraction of input energy emitted in
H$\alpha$ that we assume equal to 0.3 (Osterbrock 1987). The result is
$L_{H\alpha}=5\times 10^{-3}L_X$. This is in agreement (within a
factor of 2) with our estimate of the fraction $L_X$ converted to
H$\alpha$ photons and means that the incident X-ray flux would be
sufficient to power the observed H$\alpha$ emission. The correlation
between the H$\alpha$ EW and the veiling factor ($1-f$) finds a
natural explanation in this scenario.


\begin{figure}
\begin{center}
\centerline{\psfig{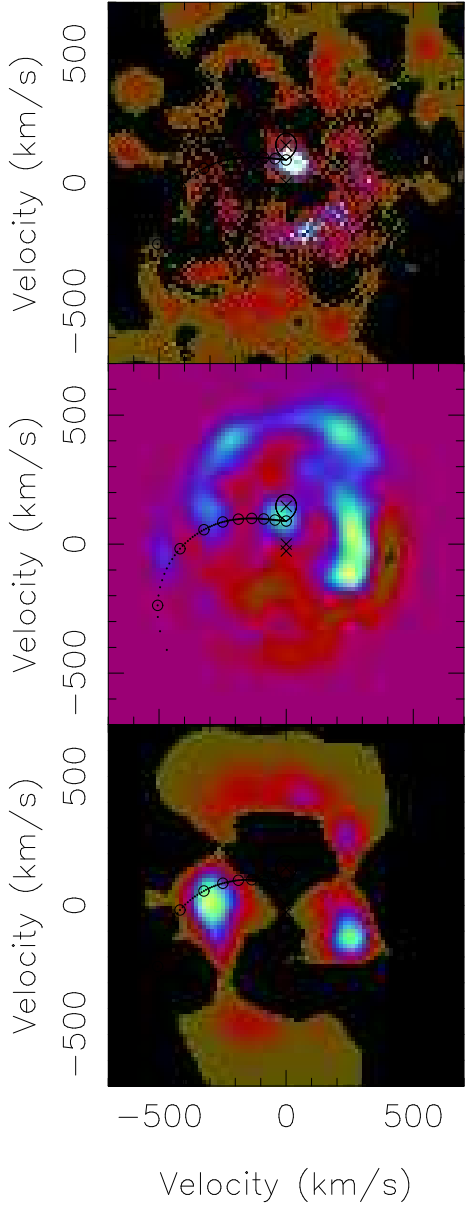}}
\caption{Doppler maps of the H$\alpha$, HeI $\lambda$5876 and HeI $\lambda$6678 lines (from top to bottom) with the axisymmetric part of the emission component subtracted. Plot symbols are the same as in fig~\ref{doppler}.}
\label{dopsub}
\end{center}
\end{figure}
 
\subsection{He line tomography} 

HeI $\lambda$5876 and HeI $\lambda$6678 Doppler maps are the first to
be obtained for a neutron star system. HeI maps, as for H$\alpha$,
show a clear accretion disc pattern, with a hint of a hot-spot
component, particularly intense in HeI $\lambda$6678. It is also
apparent from the Doppler images that the disc extends to higher
velocities in the HeI lines than in H$\alpha$, with approximate radii
at $\sim$300 km s$^{-1}$ and $\sim$200 km s$^{-1}$ respectively. This
indicates that HeI emission arises from the inner regions of the disc,
whereas H$\alpha$ emission is more concentrated in the outer disc
regions. This is as expected for viscously heated discs, where
T$_{eff}$ decreases outwards.

Also for these lines, we subtracted the disc contribution to each map by
simulating an axisymmetric image (Fig. \ref{dopsub}). After the
subtraction, both HeI maps show emission in a region consistent with
the theoretical position of the hot-spot, but only the HeI
$\lambda$5876 map shows emission coming from the secondary. These
results put in a different perspective the earlier suggestions of a
parallel with SW Sex systems. In particular, since the line energy
increases from H$\alpha$ to HeI (from 13.6 eV to 24.6 eV), it suggests
that the hot spot temperature is high enough to emit fewer H$\alpha$
photons but more He photons. Another indication of the
different temperature of the emitting region comes from the measure of
the position along the $V_y$ axis of the emission arising from the
companion in the H$\alpha$ and HeI $\lambda$5876 maps after the
subtraction of the disc contribution:\newline

H$\alpha$: $(V_y)=(97.7\pm11.8)\,{\rm km \ s^{-1}}$

HeI $\lambda$5876: $(V_y)=(122.8\pm11.8)\,{\rm km \ s^{-1}}$\newline

\noindent In a simple model where the accretion disc absorbs lower
energy photons more efficiently, we would expect lower excitation
lines to be formed higher up in the Roche lobe and show higher
velocities. A similar behaviour has been found in IP Peg by Harlaftis
(1999). This scenario, however, cannot explain our different observed
velocities. In fact, as we can see from our measurements, there is an
indication that the centroid of HeI $\lambda$5876 is more shifted
along the $V_y$ direction than that of H$\alpha$. In any case, we can
say that there is a region around $\sim$ 80 km s$^{-1}$ from L$_1$
that may mainly be clear of emission due to the shadow of the disc,
while the portion of the surface of the companion star which is
directly irradiated by the neutron star shows clear H$\alpha$ and HeI
emission.

This might indicate that the H$\alpha$ emission from the secondary
comes mainly from a region near the L$_1$ point, where the
radiation coming from the compact object is mostly occulted by the
disc. Accordingly, the HeI $\lambda$5876 emission might come from a
higher velocity region such as the portion of the surface of the
companion star which is directly irradiated (and hence hotter) by the
neutron star (i.e. it is not occulted by the disc).

\section{Conclusions}

Cen X-4 is the brightest (and nearest) quiescent SXT in the optical,
leading to its being extensively studied in recent years. It is one of
only a handful of SXTs for which H$\alpha$ Doppler tomography has been
obtained (in quiescence). Here we presented some refinements of the
companion's characteristics, classifying it as an (undermassive) K3-7
V and improving on the system parameters. We produced the first light
curve of the companion star contribution as a function of orbital
phase, confirming the shape and the model of optical and IR light
curves of Cen X-4 (McClintock \& Remillard 1990; Shahbaz et al. 1993)
and that the companion (almost) fills its Roche lobe.

Doppler tomography represents an ideal tool to test and study the
kinematics and geometry of the emission line regions of quiescent
SXTs. Our H$\alpha$ Doppler map shows significant emission at the
L$_1$ point but no emission at the predicted location of a hot
spot. The entire observed H$\alpha$ emission arising from the
companion seems to be powered by the quiescent X-ray luminosity of Cen
X-4 (Campana et al. 2004), suggesting that the companion is irradiated
by the compact object.

The H$\alpha$ Doppler map is not perfectly symmetric but shows
enhanced emission at negative $y$ velocities (see Fig. \ref{doppler}),
possibly indicating superhumps caused by eccentric precessing discs. A
similar pattern is also visible in the H$\alpha$ Doppler map of Cen
X-4 obtained by Torres et al. (2002). HeI $\lambda$5876 and
$\lambda$6678 maps add further detail to this picture. In particular,
a hot spot component appears with a centroid at HeI $\lambda$5876 that
is more shifted along the $V_y$ direction than that of
H$\alpha$. These are indications for a temperature variation, with a
hot spot able to emit higher energy lines and a cooler star emitting
H$\alpha$ and HeI $\lambda$5876 but not HeI $\lambda$6678. This can be
obtained through irradiation of the disc and hot spot region and
shielding of the L$_1$ point.



\end{document}